\documentclass[prd,
preprint,
tightenlines,
superscriptaddress,
nofootinbib,
eqsecnum,amsfonts,amsmath]{revtex4}
\usepackage{bm}
\usepackage{latexsym}
\usepackage[dvips]{graphics,color}

\begin{document}
\title{A general solution for classical sequential growth dynamics of Causal Sets}

\author{Madhavan Varadarajan}\email{madhavan@rri.res.in} 
\affiliation{Raman Research
Institute, Bangalore 560 080, India}
\author{David Rideout}\email{d.rideout@imperial.ac.uk}
\affiliation{Blackett Laboratory, Imperial College, London SW7 2AZ, UK}

\begin{abstract}
A classical precursor to a full quantum dynamics for causal sets has been formulated
in terms of a stochastic sequential growth process in which the elements of the
causal set arise in a sort of accretion process.
The transition probabilities of the Markov growth process satisfy
certain physical requirements of causality and general covariance, and the generic solution with all transition probabilities non-zero has been found.
Here we remove the assumption of non-zero probabilities, 
define a reasonable extension of the physical requirements to cover the case of vanishing probabilities, 
and find the completely general solution to these physical conditions.
The resulting family of growth processes
has an interesting
structure reminiscent of an ``infinite tower of turtles'' cosmology.

\end{abstract}
\preprint{gr-qc/0504066}
\maketitle

\widetext
\baselineskip22pt

\section{Introduction}
The causal set approach to quantum gravity posits that the deep structure of
spacetime is a locally finite partially ordered set \cite{luca}. One of the 
key open questions is a formulation of a quantum dynamics for causal sets.
As a preliminary step towards such a formulation,
one can
define a {\em classical} stochastic dynamics for causal sets in terms of a 
sequential growth process in discrete stages, each of which involves the 
addition of a new element to a causal set obtained from the previous stage.
In this context, the dynamical law is an assignation of  
probabilities to each such transition from every finite causal set to its 
possible `children' in accordance with certain physical principles 
inspired by the continuum notions of general covariance and causality
\cite{RS}.\footnote{The dynamics is classical in that no allowance is made
  for quantum interference between 
possible distinct transitions from any causal set to its children. 
A quantum dynamics would be expressed in terms of a \emph{quantum measure}, or \emph{decoherence functional},
which generalizes the notion of probability measure to allow for interference
of distinct possibilities. \cite{quantummeasure}}
Ref.\ \cite{RS} finds the most general solution for the transition 
probabilities subject to these principles {\em and} the additional 
assumption that none of the transition probabilities vanish. Thus the
solution is generic but not the most general one.

The causal sets which arise from the generic classical sequential growth models are reasonably well understood.  For example, there is significant evidence indicating that they do not produce `manifoldlike' causal sets \cite{georgiou005}.
It is of interest to know whether the picture changes significantly if we allow
vanishing transition probabilities.
The quantum theory of causal sets is expected to arise from a
decoherence functional (or quantum measure) defined on sets of histories
(causal sets).  In some appropriate limit (say after coarse graining to
achieve decoherence), one expects to get probabilities which obey a Kolmogorov
sum rule, and there is no reason to expect that none of these will vanish.
Thus it is important to know if there is any drastic effect which arises
from the case of vanishing probabilities.

In this work we extend the considerations of ref.\ \cite{RS} to the general case 
in which the transition probabilities are required merely to be non-negative rather than
positive. 
We will see that the general dynamics which results can be regarded as a
sequence of different copies of the generic dynamics described in ref.\ \cite{RS},
each being a `turtle' in an infinite (temporal) tower of
turtles\footnote{This is in reference to a popular legend about and old woman
  who, at the end of a lecture by a famous scientist, 
  attempts to argue that the Earth is really flat and rests on the back of an
  infinite tower of turtles.}
.  This has
been investigated earlier by Joohan Lee \cite{joohan}.
In our treatment we assume some familiarity with the terminology 
and proofs of ref.\ \cite{RS}.

The outline of the paper is as follows.
We extend the physical principles of discrete general covariance and Bell 
causality to the case of vanishing transition probabilities
in section II. These principles, in conjunction with the Markovian and 
internally temporal nature of the growth process \cite{RS}, restrict
the dynamical law in certain important ways. We derive these restrictions 
in section III. In section IV we show that these restrictions taken together
allow an explicit characterization of the most general classical dynamics
describing a growth process consistent with the physical principles mentioned 
above. Section V contains our conclusions, and a few useful lemmas are proved
in the Appendix.

\section{Physical requirements on the dynamics}
\label{phys_req}

As in ref.\ \cite{RS}, consider ${\cal P}$, the poset 
of all  (isomorphism equivalence classes of) finite causal sets (causets) wherein if a
causet can be 
formed by accreting a single element to a second causet, then the former (the
`child') follows the latter (the `parent') in ${\cal P}$ and the relation 
between the causets is a link (a relation not implied by transitivity).
 A sequential growth process corresponds to a path (i.e.\ a series of transitions from
one causet to another)
in $\cal P$, starting from the empty causet.
Recall that a link may correspond to more than one distinct 
transition (the number of distinct transitions are the number of inequivalent
embeddings of the parent as a partial stem of the child, where two embeddings
are equivalent if related by an automorphism of the child; a partial stem is
a sub-causet which contains its own past).
A \emph{dynamical law}
is defined to be an assignation
of transition probabilities (i.e.\ real numbers in [0,1]) to each such distinct
transition for every link
of $\cal P$. 
We shall require that 
the dynamical law be  consistent with 
the principles of  
general covariance and Bell causality 
as well as the Markov sum rule defined below in sections A--C.
As noted in ref.\ \cite{RS}, 
the dynamics, by virtue of its formulation as
a sequential growth process, automatically incorporates 
the property of internal temporality (which simply means that no new
element can be born to the past of an existing element of any parent).

As in ref.\ \cite{RS}, we set the probability $q_0$ of forming the single element
causet (a 1-chain) to unity.

\subsection{General Covariance}
A dynamical law is defined to be generally covariant iff,\\
\noindent (i) The transition probabilities for distinct transitions 
associated with the same link in ${\cal P}$ are identical. \\
\noindent (ii) If $\gamma$ is any path through $\cal P$ 
which originates at the empty set and terminates at a causet $C$, the 
product of transition probabilities along its links
is the same as for any other path from the empty set to $C$.  

For any generally covariant dynamics, we 
shall refer to the product of transition 
probabilities 
along the links of a path connecting the empty set to a causet 
$C$ as the 
{\em specific
probability of formation} of $C$.
\footnote{This is contrast to the {\em total}
probability of formation of $C$ defined in ref.\ \cite{RS}.  The latter is 
obtained by multiplying the specific probability by the number of
inequivalent natural labellings of $C$ \cite{RS}.  Note that even the total
probability of formation lacks covariant meaning, in that it refers to the
probability of forming a particular finite causet after a specific stage of
the growth process.  Physically meaningful probabilities can be extracted
from the sequential growth dynamics by the measure it assigns to cylinder
sets of unlabeled causets, as described in ref.\ \cite{RSmeasure}.
}

For any assignation of transition 
probabilities consistent with general covariance, we define a 
{\em virtual} causet as one whose {\em specific probability of formation}
vanishes. Thus any path from the empty set to a virtual causet contains 
at least one
link with zero transition probability. All causets which are not 
virtual are called {\em real}. Virtual causets (and hence, their descendants)
are never formed in the growth process. This is the reason that our definitions
of Bell causality and the Markov sum rule below pertain only to real parents.
Since only specific probabilities of formation are of interest, two dynamical 
laws which generate the same set of specific probabilities of formation will
be referred to as \emph{equivalent}.
We restrict our considerations in the remainder of this section to 
generally covariant dynamics.

\subsection{The Markov Sum Rule} We impose the same requirement as in ref.\
\cite{RS}, 
except that we demand it only of real parents. Thus, we require that the sum
of the full set of transitions issuing from a given {\em real} causet is unity.
(The full set of transitions constitutes one for each choice of partial stem
of the parent.  The coefficients in the sum rule of ref.\ \cite{RS} arise 
when multiple partial stems result in the same child causet.)

\subsection{Bell Causality}
As mentioned above, Bell  causality 
is only defined for {\em real}
parents. Let $C$ be a real parent and $C_1$ and $C_2$ be two of its 
children. Let $B$ be the union of the precursor sets 
for the two transitions.  (Recall that a precursor set is the past of the new 
element whose introduction forms the child causet $C_1$ or $C_2$.)
Clearly there is a path in ${\cal P}$ starting from the empty set,
ending at $C$ and passing through $B$. Since $C$ is real so is $B$. 

Let $B_1$ and $B_2$ be the causets defined by adding an element to the future of 
the corresponding precursor sets in $B$  
and let $P(C\rightarrow C_i),P(B\rightarrow B_i)$ be the transition 
probabilities for the transitions $C\rightarrow C_i, B\rightarrow B_i$, $i=1,2$.
In ref.\ \cite{RS}, Bell causality was formulated as
\begin{equation}
\frac{P(C\rightarrow C_1)}{P(C\rightarrow C_2)} = \frac{P(B\rightarrow B_1)}{P(B\rightarrow B_2)} .
\label{bcrs}
\end{equation}
This equation was meant to capture the idea that events occurring in some part of a causet 
should be influenced only by the portion of the causal set lying to their past.
Equation (\ref{bcrs}) is only sensible when all transition probabilities are non-vanishing as in ref.\ \cite{RS}.
We seek a generalization of this equation to the case of vanishing transition
probabilities (i.e.\ when
one or more of the children in (\ref{bcrs}) are virtual).

Let $P(C\rightarrow C^{\prime})=0$ for some transition from a real parent $C$ to its child $C^{\prime}$.
Then a natural condition inspired by the idea alluded to 
just after equation (\ref{bcrs}) 
would be to
forbid {\em all} transitions from real parents which involve the same (isomorphism class of) precursor 
set as in $C\rightarrow C^{\prime}$.
Since the transition from the empty set to the 1-chain has probability $q_0=1$, it would follow from such 
a condition that no antichain to antichain transition could be virtual. 
As we shall 
see in 6(b) of the proof of Lemma 2 in section  III, 
the general solution to such a dynamics is that of ref.\ \cite{RS} in which the
$t_k$ 
can be zero.

Our aim is to provide as general a dynamical law as
possible, following the spirit of the conditions imposed in ref.\ \cite{RS}. 
In particular, we would like 
to allow for the vanishing of any of the transition probabilities, including those of the antichain to antichain transitions.
A natural set of conditions which allows this and serves as a reasonable generalization of equation (\ref{bcrs})
may be arrived at by the following qualitative discussion.

The formulation of (\ref{bcrs}) may be thought to involve two distinct ingredients: \\
\noindent (a) The idea that there is a propensity for a transition to occur depending solely on the 
nature of the transition, i.e.\ the precursor set involved. This is incorporated in (\ref{bcrs}) by requiring that 
the transitions $C\rightarrow C_i, B\rightarrow B_i$ have the same precursor sets for each of $i=1,2$. \\
\noindent (b) The implication of the Markov sum rule  that the net probability of formation of all possible children from a real 
parent is fixed and equal to unity. This forces (\ref{bcrs}) to be an equality of {\em ratios} of probabilities
rather than probabilities themselves. 

Viewed in terms of (a) and (b), a transition probability could vanish due to two distinct reasons:
(1) the transition is intrinsically forbidden so that all transitions involving the same 
precursor set are also  virtual or (2) there are so many competing siblings that they
``take away the entire available probability'' and drive the transition
probability for the transition in question to zero.
The consequences of this viewpoint are as follows.
Consider, as before,  the real causet $C$ and its ancestor $B$. Every precursor set in $B$ is also a
precursor set in $C$. Hence for every birth in $B$ there is a birth in $C$, but since $C$ is larger than 
$B$, it has more children than $B$. Clearly, if $P(B\rightarrow B_1) = 0$ then $C_1$ must also
be virtual since $C_1$ has even more competing siblings than $B_1$. On the other hand, if 
$P(C\rightarrow C_1) \neq 0$, then $B_1$ must also be real, since $B_1$ has even fewer siblings than
$C_1$. The relative propensity of the birth of $C_2$ with respect to that of $C_1$ is then well defined
as the ratio of the two transition probabilities,
and may be taken to quantify the relative propensity of the birth $B_2$ with
respect to that of $B_1$.
In the case that both $C_1$ and $C_2$ are virtual, the relative propensity of their births
is ill defined. The vanishing of $P(C\rightarrow C_i), i=1,2$  may be because $C_1$ and $C_2$ 
have too many competing siblings. The corresponding 
children $B_1,B_2$ of $B$ have fewer competing siblings and hence it is possible that either or both 
these children are real. 

As a result of this qualitative discussion, we formulate the Bell causality condition in terms
of $C,C_i,B,B_i,\;i=1,2$ (which have been defined just
before equation (\ref{bcrs})) as follows:

\noindent (i) If all four transition probabilities are non-vanishing, Bell causality 
is defined by (\ref{bcrs}).

\noindent (ii) If $P(B\rightarrow B_1) = 0$, then
$P(C\rightarrow C_1) = 0$. If  $P(B\rightarrow B_2) = 0$
then $P(C\rightarrow C_2) = 0$. 

\noindent (iii) If $P(C\rightarrow C_1) = 0$, 
$P(C\rightarrow C_2) \neq  0$, then 
$P(B\rightarrow B_1) = 0$, $P(B\rightarrow B_2) \neq 0$

\noindent (iv) If $P(C\rightarrow C_1) = 0$ and $P(C\rightarrow C_2) = 0$
then nothing can be inferred about  $P(B\rightarrow B_1)$ and 
$P(B\rightarrow B_2)$.

Note that if the transition probabilities vanished only because the transition in question was intrinsically forbidden,
we would obtain that $P(C\rightarrow C_i) = 0$ iff $P(B\rightarrow B_i) = 0,\;\;i=1,2$ which would in turn imply
(i)--(iii) above and a {\em stronger} condition than (iv).  
In the remainder of this work we  use (i)--(iv) as our definition of Bell causality.  In addition we freely make use of the fact that the Bell causality
conditions (i)--(iv) hold when $B$ is replaced by any subcauset of $C$ which contains
the union of the precursors of the two transitions as a partial stem.
That this is so can easily be verified for any
assignment of transition probabilities consistent with general covariance as defined above.

\section{ Some Implications of the Physical Requirements}

A few key consequences on sequential growth dynamics, of general covariance, Bell causality and the Markov sum rule,
are derived in this section.

\vspace{5mm}

\noindent{\bf Lemma 1}: Let the $j$-antichain to $j+1$-antichain 
transition probability $q_j$ be such that $q_j >0$ for $j=1..n$.
Then for $J \leq n$, the gregarious transition from any real parent of cardinality 
$J$ has probability $q_J$.  (Recall the gregarious transition is that in which the new element arises unrelated to any of the elements of the parent causet.)

\noindent {\bf Proof}:  Clearly, we need consider parents which are not 
antichains. Let such a real parent  be $A_0$ with cardinality $J$.
\begin{figure}[htbp]
\center
\scalebox{.85}{\includegraphics{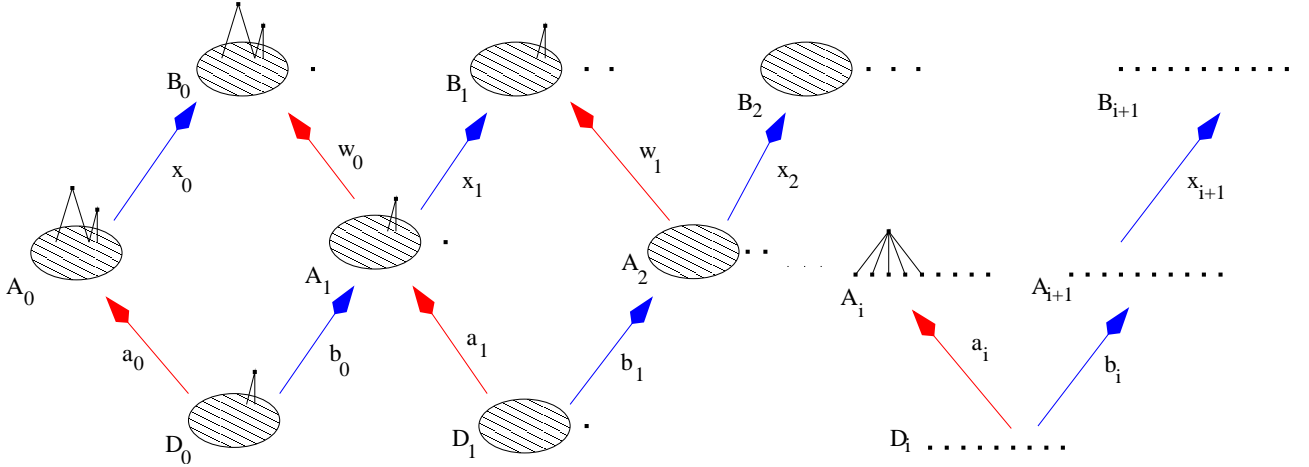}}
\caption{All ``gregarious child'' transitions have probability $q_J$}
\label{greg_trans}
\end{figure}
Refer to Fig.\ \ref{greg_trans}.  
The causets $B_j$ are the gregarious children of the $A_j$.
The $D_j$ are
parents of the $A_j$, as shown schematically in the diagram.  The transition
probabilities between various pairs of causets are as labeled.
Since $A_0$ is real, general covariance (referred to henceforth as g.c.) implies that $a_0\neq 0$.
Suppose $x_0 \neq 0$. Then $B_0$ is real and g.c.\ implies that 
$b_0,w_0 \neq 0$. 
Now employ Bell causality (henceforth referred to as b.c.) to compare
the transition probabilities $w_0$ and $x_1$ with $a_0$ and $b_0$ respectively,
where a disconnected element of $A_1$ acts as a spectator.  (Recall a spectator is an element of the parent causet which does not lie in the precursor set of either of the transitions in question.)
Since both $a_0$
and $b_0$ are non-vanishing, along with $w_0$, b.c.\ (iii) forces $x_1$ to be non-zero as well.
Thus we may use b.c.\ (i) to prove that $x_0=x_1$ as in ref.\ \cite{RS} (simply
consider the b.c.\ (i) condition $\frac{b_0}{a_0} = \frac{x_1}{w_0}$ along with
the g.c.\ condition $a_0x_0 = b_0w_0$).
%
This means that $B_1$ is real.
Since $D_1$ and $A_1$ must be real, and $a_1\neq 0$, we can repeat the
argument with $A_1$ in the place of $A_0$, and
proceed rightwards across the figure.

Clearly, as in ref.\ \cite{RS}, the process terminates for $i$ such that:
$A_i$ has only one maximal element with non empty past, $A_{i+1}$ is the 
$J$ antichain, $B_{i+1}$ is the $J+1$ antichain and $D_i$ is the 
$J-1$ antichain. Then b.c.\ and g.c.\ imply that $x_0 = q_J$.

Above we assumed that $x_0 \neq 0$.  To complete the proof, we show by
contradiction that this is so.  Thus suppose $x_0=0$.  Then

\noindent (a) Suppose $A_1$ is real. 
$\Rightarrow b_0\neq 0$. Then g.c.\ implies that $w_0=0$.
Then b.c.\ (iii) implies that $x_1=0$. (Again, the disconnected element of $A_1$
acts as the spectator.  Since $w_0=0$, only $x_1=0$ is consistent with 
b.c.\ (iii).)
Further since $A_1$ is real,
$a_1\neq 0$ and we can repeat the argument with $A_1$ as the new $A_0$, asking
now if $A_2$, which is the new $A_1$, is real.  In this manner we can
progress rightwards across the figure (assuming the $A_j$ are real, c.f.\ case
(b) below)

\noindent (b) Suppose $A_1$ is virtual. Since $A_0$ is real, we have that 
$D_0$ is real and hence $b_0=0$. Then we can not proceed rightwards across the figure
 since b.c.\ does not apply for
virtual parents. But since $D_0$ is real we can
choose $D_0$ to be our new $A_0$, with $b_0=0$ now taking the role of $x_0=0$,
and repeat the argument (by asking again if the new $A_1$ is real).
Note that this recursive argument must eventually end with case (a).
As a `worst case', this recursion will eventually arrive 
at the 2-chain for $A_0$, whose only corresponding $A_1$ is the 2-antichain,
which we know is real even for $n=1$ (since $q_1>0$).

This procedure will 
terminate with some final choice of real $A_0$ with cardinality $K$,
$2\leq K \leq J$, such that $A_0$ has only one maximal element with a nonempty
past, $D_0$ is the $K-1$ antichain, $A_1$ is the $K$ antichain, and $B_1$
is the $K+1$ antichain.  Then $x_1$ will be $q_K$, $K \leq J \leq n$, which is
non-zero.  But from (a) above $x_1=0$, which is a contradiction.
%
This completes the proof of the Lemma.

\vspace{5mm}

\noindent {\bf Lemma 2}:
 Let the antichain to antichain transition
probabilities $q_1,..,q_n$ all be $>0$. 
Then the most general sequential growth 
dynamics to stage $n$ furnishes probabilities of formation of 
causets of cardinality $\leq n+1$  in accordance with equation (12) of
ref.\ \cite{RS},
\begin{displaymath}
  \alpha_n = \frac
             {\sum_{l=m}^{\varpi} {\varpi-m \choose \varpi-l} t_l}
	     {\sum_{j=0}^n {n \choose j} t_j}
\end{displaymath}
in which the coupling constants $t_k, k=1..n$ are such that 
$t_k \geq 0$, or equivalently in accordance with equation (7) of ref.\ \cite{RS}, 
\begin{displaymath}
  \alpha_n = 
	\sum_{k=0}^m (-)^k {m \choose k} \frac{q_n}{q_{\varpi-k}}
\end{displaymath}
with 
\begin{displaymath}
\sum_{l=0}^n (-)^{n-l} {n \choose l} \frac{1}{q_l} \geq 0 \;.
\end{displaymath}

\noindent {\bf Proof}: Consider a sequential growth dynamics consistent with
 b.c., g.c.\ and the Markov sum rule. 
It is straightforward to repeat the 
considerations of section IV of ref.\ \cite{RS} {\em for real parents}.  Here we
briefly repeat the arguments, taking into account the possibility of vanishing
transition probabilities.

\vspace{3mm}

\noindent (1) Lemma 1 holds so that any gregarious transition from a $J$ element
real parent has  transition probability $q_J, J\leq n$. (Lemma 1 is the 
analog of Lemma 2 in ref.\ \cite{RS}).

\vspace{3mm}

\noindent (2) {\em Claim}: The analog of Lemma 3 in ref.\ \cite{RS} holds for real parents. Thus each
transition probability  
$\alpha_m$ of stage $m\leq n$ from a {\em real}
parent has the form 
\begin{equation}
\alpha_m = q_m\sum_{i=0}^{m} \frac{\lambda_i}{q_i} 
\label{3}
\end{equation}
where
$\lambda_i$ are integers {\em only} depending on the transition in question.\\
\noindent {\em Proof by induction}:
Equation (\ref{3}) is easily verified for stages 0 and 1. In particular
stage 1 always has 
one real parent and the above form for the $\alpha_1$ holds. Assume (\ref{3}) holds
for stage $k-1 < n$. Consider a bold transition probability $\beta_k$ of stage $k$
from some real parent $C$.  (Recall the timid transition is the one in which
the new element arises to the future of the entire parent causet.  A bold
transition is any save the timid transition.)
 For any such  causet, there exists a real 
parent $B$ at stage $k-1$ such that the bold and
gregarious transitions from $C$ are in b.c.\ with appropriate ones from $B$. $B$ can be
constructed by removing from $C$ a maximal element which is not in the 
precursor set for the bold transition. It is easy to see that g.c.\ implies
that $B$ is real if $C$ is real. Then b.c.\ (i \& iii) gives 
$\frac{\beta_{k}}{q_k}= \frac{\alpha_{k-1}}{q_{k-1}} \implies \beta_k = q_k\sum_{i=0}^{k-1}\frac{\lambda_i}{q_i}$
irrespective of whether the bold child of $C$ is virtual or not. The Markov sum rule
ensures that the timid child transition probability, $\gamma_k$ is given by
$\gamma_k = 1-\sum_{j}\beta_{kj}= 1 - q_k \sum_{i=0}^{k-1}\sum_{j}\frac{\lambda_{ij}}{q_i}$ 
where $j$ labels the possible bold transitions. As in ref.\ \cite{RS} this expression
can be put in the form (\ref{3}) by setting $\lambda_i = -\sum_{j}\lambda_{ij},i <k$
and $\lambda_k =1$.

\vspace{3mm}

\noindent (3) In transitive percolation {\em all} causets are real, save the
special cases when $p=0$ or $p=1$.  From ref.\ \cite{RS} we know that, for transitive
percolation, $q_n = q^n$, where $q=1-p$.  The case of $p=1$ is disallowed by
assumption, because in that case all $q$'s vanish.  The case $p=0$ makes all
$q_n=1$, yielding an infinite antichain with probability 1.  In the general
case 
$\{\lambda_i\}$ for any transition from a real parent can be obtained from a comparison of 
$\alpha_k =q_k\sum_{i=0}^{k} \frac{\lambda_i}{q_i}$ ($k\leq n$) with $\alpha_k$ for
transitive percolation just as in ref.\ \cite{RS}  and we get the same answers as in ref.\ \cite{RS} namely
\begin{equation}
\alpha_k = \sum_{i=0}^m (-)^i{m \choose i}\frac{q_k}{q_{\varpi -i}}
\label{7}
\end{equation}
where $\varpi$ is the cardinality of the precursor set for the transition
and $m$ is the number of its maximal elements.  (Note that this formula also
gives probabilities consistent with transitive percolation when $p=0$,
namely that $\alpha_k=1$ when $m=0$, and 0 otherwise.)

\vspace{3mm}

\noindent (4) {\em Claim}:  In order that for all transitions from real parents the transition probabilities 
$\alpha_k \in [0,1]$, it suffices that 
each timid transition probability for every real parent is $\geq 0$, which is in turn guaranteed 
if the timid transition probability from the $k$-antichain is $\geq 0$ for all $k\leq n$.
\footnote{Note that the   $k$-antichain, $k\leq n+1$ is real since there is a path from the 
empty set to the $k$-antichain comprising only of antichains and the transition probability for the 
$i$th link is $q_i >0$.}\\
\noindent {\em Proof}: The proof of this statement is identical to that 
of ref.\ \cite{RS} restricted to real parents, save that each reference to a
non-gregarious transition probability being positive is replaced by the
statement that it is non-negative.
In order to use our definition of b.c.\ in the proof, it suffices 
to note (as in (2) above) that for every bold transition from a real parent $C$ at stage $k$ ($2\leq k\leq n$), 
there is a real parent $B$ of $C$ which contains the precursor set for the
transition.  Since $q_k$ and $q_{k+1}$ are both non-zero, our b.c.\
(i--iii) is equivalent to the b.c.\ formula used in the proof, where now the
$\alpha$ can vanish.

\noindent (5) From (4) it follows, in an identical manner to the considerations of ref.\ \cite{RS}, that 
the transition probabilities $\alpha_k, k\leq n$ from real parents of size $k$ are of the form
\begin{equation}
\alpha_k = q_k\sum^{\varpi}_{l=m}{\varpi -m \choose \varpi -l}t_l
\label{12}
\end{equation}
with $q_k$ expressible in terms of $t_k$ as 
\begin{equation}
\frac{1}{q_k} = \sum^{k}_{l=0} {k \choose l}t_l .
\label{qt}
\end{equation}
As in ref.\ \cite{RS} the `coupling constants' $t_i$ can be freely chosen subject to
the conditions $t_0=1, t_i\geq 0, 1\leq i\leq n$.

\vspace{3mm}

\noindent (6) {\em Claim}: The transition probabilities for transitions from real parents
given by the formula (\ref{12})--(\ref{qt}) with $t_j\geq 0, n\geq j\geq 1, t_0=1$,
satisfy the physical requirements. \\
\noindent {\em Proof}: The proof is obtained simply by restricting the proof
of section IV D of ref.\ \cite{RS}
to real parents. We briefly describe how the results of ref.\ \cite{RS} apply here.

\noindent (a) General Covariance:
Assign transition probabilities to {\em all} links (till stage $n$) in
$\cal P$
according to the formula (\ref{7}) or (\ref{12}).\footnote{Note that these transition probabilities are non-negative and
identical for all transitions corresponding to the same link in ${\cal P}$.
Hence such an assignation is consistent and defines a dynamical law.}
It is easily verified that 
for any path from the empty set to a causet $C$ of cardinality $|C|\leq n+1$,  
the product of transition probabilities, apart from the overall factor $\prod_{j=0}^{|C|-1}q_j$,
is a product of factors which are in one to one correspondence with elements of $C$ such that the factor 
corresponding to the element $x$ only depends on the past of $x$ in $C$. Hence this product is
path independent. 
If this product  $\neq 0$, then $C$ is real, otherwise it is virtual.

\noindent (b) Bell Causality:
Restrict equation (\ref{7}) to real parents. Then the transition probability for any birth depends,
apart from a factor of $q_k$, only on the precursor set for the transition. As in ref.\ \cite{RS}, this
implies that b.c.\ (i) holds. It also implies that if any transition probability vanishes {\em all}
births involving the same precursor set are virtual. Thus the dynamics satisfies a stronger version
of b.c.\ than we require (see the comment at the end of section
\ref{phys_req}).  In this regard, note that b.c.\ (iv) did not come in to our derivation of
the general dynamical law.  Thus b.c.\ (i--iii), in the presence of general covariance and
the Markov sum rule, implies a stronger causality condition that that
expressed in b.c.\ (iv).  (Though the situation changes when we allow the
gregarious transition probabilities to vanish, and our weaker notion of b.c.\
becomes important in that context.)

\noindent (c) Markov Sum Rule: The proof in ref.\ \cite{RS} goes through without any change. The proof applies
to {\em any} parents and we can simply restrict the proof
to real parents.

\vspace{5mm}

\noindent {\bf Lemma 3}: Let $n$ be the smallest number for which the 
$n$-antichain to $n+1$-antichain transition probability $q_n$ vanishes.
Then: (a) There are no real gregarious children at stage $j$, $j\geq n$.
(b) At stage $n$ the only real children are timid children.

\noindent {\bf Proof}:

\noindent (a) Again consider Fig.\ \ref{greg_trans}.
Let $A_0$ be a real parent of cardinality $\geq n$ and assume that $B_0$ is
its real gregarious child. Hence $a_0, x_0 \neq 0$.  As in the proof of Lemma
1, using g.c.\ and b.c.\ (iii), 
it is easy to see that all the causets of Fig.\ \ref{greg_trans} must be real.
But these 
causets include the $j+1$ antichain, $j\geq n$.  Since $q_n=0$, there is 
a path (solely consisting of antichains) from the empty causet to the $j+1$ 
antichain which has specific probability $=0$. Hence the $j+1$ antichain 
is virtual. This yields the desired contradiction. Hence $B_0$ cannot be real.

\noindent (b) Let $C$ be a real parent (of cardinality $n$) at stage $n$.
Consider the b.c.\ relation involving any bold child $C_b$ of $C$ and the
 gregarious child $C_g$  of $C$. Let the precursor of $C\rightarrow C_b$
be $C^{pre}$ and the relevant (timid and gregarious) children of $C^{pre}$
be $C^{pre}_{t}, C^{pre}_g$.

Clearly, there is a path in ${\cal P}$ from the empty set through $C^{pre}$ to $C$.
Hence, since $C$ is real, so is 
$C^{pre}$. Let us further  assume that 
$C_b$ is real.  The same reasoning implies that $C^{pre}_t$ is real. Then we have that  \\
\noindent $P(C^{pre}\rightarrow C^{pre}_t) \neq 0$, 
$P(C\rightarrow C_b) \neq 0$.  \\
\noindent Let the cardinality of $C^{pre}$ be $K$ ($K<n$ since
$C_b$ is bold). 
Then Lemma 1 implies that  
$P(C^{pre}\rightarrow C^{pre}_g) =q_K\neq 0$. Also (a) above together with 
the fact that $C$ is real,  implies that $P(C\rightarrow C_g) = 0$. 

But this assignment of
transition probabilities is in contradiction with b.c.\ (iii). Hence $C_b$ cannot
be real. (Note that with $C_b$ virtual, b.c.\ (iv) is satisfied.)  Hence the only real children produced at stage $n$ are timid children.

\noindent{\em Implications of Lemma 3}: To describe the implications of Lemma 3
and for our subsequent considerations, it is useful to define the notion of 
a $C$-{\em timid} causet as follows. We shall say that 
a causet $C^{\prime}$ is {\em timid with 
respect to a causet $C$}, or that $C^{\prime}$ is {\em $C$-timid}, if 
$C^{\prime}\supset C$ and every element
in $C^{\prime}\setminus C$ is to the future of every element of $C$.

Now, let
$n$ be the minimum stage at which $q_n=0$ and let there be $P$ real (non-isomorphic)
parents at stage $n$ formed as a result of sequential growth. Denote these as 
$C_{j,n}$, $j= 1,..,P$. By Lemma 3, only the timid transition is allowed at stage $n$.
This means that at any subsequent stage the new element added must be to the 
future of the entirety of $C_{j,n}$.  If not, a real growth process could be envisaged such that
the child at the $n$th stage was not timid.  
{\em Thus, every real causet formed as a result of 
growth after stage $n$ is $C_{j,n}$-timid for some unique $j$} (the uniqueness follows from Lemma A3 
in the Appendix).


\section{The General Solution}

In this section we derive the most general solution to the dynamics which
satisfies the physical requirements of section II. The derivation uses the 
results of section III and Lemmas A1, A2 and A3 proved in the Appendix.
We shall present our derivation in the form of two Lemmas 4A and 4B and a Remark. 
Their import is as follows.

Let ${\cal P}_{C_{j,n}}$ be the subposet of ${\cal P}$ which contains $C_{j,n}$ and 
all $C_{j,n}$-timid causets for a fixed $j$ (we have defined $C_{j,n}$ at the end of the previous section above). 
We define a dynamical law {\em relative} to 
${\cal P}_{C_{j,n}}$ to be an assignation of probabilities to links in ${\cal P}_{C_{j,n}}$. Such 
a dynamical law will be said to satisfy the physical conditions of section II {\em relative}
to ${\cal P}_{C_{j,n}}$ iff \\
\noindent (a) the transition probabilities for distinct transitions 
associated with the same link in ${\cal P}_{C_{j,n}}$ are identical and the product of 
transition probabilities along any path in ${\cal P}_{C_{j,n}}$ starting from $C_{j,n}$ to a causet
$C$ in ${\cal P}_{C_{j,n}}$ depends only on $C$.\\
\noindent (b) b.c.\ as defined in section II holds among causets in ${\cal P}_{C_{j,n}}$. \\
\noindent (c) the sum of transition probabilities for all $C_{j,n}$-timid children of any real
parent in ${\cal P}_{C_{j,n}}$ is unity, where reality is defined with respect to paths in 
 ${\cal P}_{C_{j,n}}$ starting from $C_{j,n}$. 

Lemmas 4A and 4B show that in order to find the most general dynamical law consistent 
with the conditions of section II, it suffices to find the most general dynamical law relative
to ${\cal P}_{C_{j,n}}$ which satisfies the physical principles of section II relative to 
${\cal P}_{C_{j,n}}$, for each $j$ separately. Remark 1 shows that the latter assignation of 
transition probabilities is in correspondence with a growth process from the empty set. As we shall
see, by applying Lemmas 4A, 4B and Remark 1
iteratively, we shall be able to derive the general solution to the dynamics in ${\cal P}$.

Let ${\cal S}_n$ be a dynamical law for causets till stage $n$ (i.e.\ the maximum size of children is $n+1$)
which is specified as follows. Let the transition probabilities till stage $n-1$ be assigned in
accordance with (\ref{7}) or (\ref{12}). As in the last paragraph of the previous section,
let $q_n=0$ and let $C_{j,n}, j=1..P$ be the real non-isomorphic parents of size $n$.
Further, let the timid transition from each $C_{j,n}$ occur with unit probability and every other
transition at stage $n$ with probability zero.

\vspace{3mm}

\noindent{\bf Lemma 4A}:  Let ${\cal S}$ be a dynamical law whose restriction
up to stage $n$ is ${\cal S}_n$.
If ${\cal S}$ is consistent with g.c., the Markov sum rule and b.c.\ then an equivalent dynamical law 
\footnote{See IIA for a definition of equivalent dynamical laws.} exists such that\\

\noindent (a) at stage $r\geq n$ transition probabilities for all transitions to causets which 
are not timid with respect to any $C_{j,n}$ vanish. \\
\noindent (b) g.c., the Markov sum rule and b.c.\ as defined in section II
 hold.

\noindent{\bf Proof}:  
Since the causets described in (a) are virtual by Lemma 3, Lemma A1 ensures that we may define an
equivalent dynamical law ${\cal S}^{\prime}$ by setting the transition  probabilities of (a) to zero.
(b) is trivially true for ${\cal S}^{\prime}$ 
from Lemma A1.


\vspace{5mm}

\noindent{\bf Lemma 4B}:
Let ${\cal S}$ be an assignment of transition probabilities (i.e.\ numbers in [0,1]) to links in 
${\cal P}$  such that \\
\noindent (a) ${\cal S}$ coincides with ${\cal S}_n$ to stage $n$. \\
\noindent Beyond stage $n$:\\
\noindent (b) transition probabilities for all transitions to causets which 
are not timid with respect to any $C_{j,n}$ vanish. \\
\noindent (c) ${\cal S}$ restricted to each ${\cal P}_{C_{j,n}}$ provides a dynamical law
relative to ${\cal P}_{C_{j,n}}$ which satisfies the principles of section II relative to
${\cal P}_{C_{j,n}}$ for each $j$ separately.

Then ${\cal S}$ is completely specified by (a)--(c) (i.e.\ (a)--(c) ensure that every link in ${\cal P}$ is 
assigned a unique number in $[0,1]$) and is consistent with the physical principles of section II.\\

\noindent{\bf Proof}:\\
\noindent{\em Claim 1}: ${\cal S}$ is completely specified by (a)--(c).\\
\noindent {\em Proof}: Transition probabilities from any causet of size $\leq n$ are specified by 
${\cal S}_n$. Any causet of size $>n$ is either $C_{j,n}$-timid for some $j$ or not timid
with respect to any $C_{j,n}$. 

If the latter then Lemma A2 shows that its only offspring are non-timid with
respect to any of the $C_{j,n}$ and (b)
specifies the transition probabilities.
If the former, 
then Lemma A2 ensures that it is not
$C_{k,n}$-timid for any $k\neq j$. Its children are either $C_{j,n}$-timid in which case (c)
specifies the transition probabilities or, using Lemma A2, non-timid with respect to any 
$C_{k,n}, k=1..P$ in which case (b) specifies the transition probabilities.

\vspace{3mm}

\noindent {\em Claim 2}: If (a)--(c) hold, ${\cal S}$ is generally covariant.\\
\noindent {\em Proof}: From Lemma A2 and the consistency of the dynamics of
Eqns. (\ref{7}) and (\ref{12})
with g.c., ${\cal S}_n$ is clearly consistent with g.c. Hence g.c.\ needs to be checked only for 
causets of size $> n+1$.  If the causet is not timid with respect to any $C_{j,n}$, then by 
(b) every path to it has at least one link with zero transition probability. Hence such 
paths satisfy g.c.  If the causet is $C_{j,n}$-timid, Lemma A2 shows that every path to it
is such that \\
\noindent (i) it passes through $C_{j,n}$. \\
\noindent (ii) it does not pass through any causets of cardinality $>n$ which are not timid with respect to any $C_{k,n}$, $k=1..P$,
nor through any $C_{k,n}$-timid causets, $k\neq j$ thus implying that it must pass through only
$C_{j,n}$-timid causets after stage $n$.

Since ${\cal S}_n$ is consistent with g.c., (i) and (ii) in conjunction with (c) show that paths to 
$C_{j,n}$-timid causets are consistent with g.c.

\vspace{3mm}
\noindent {\em Claim 3}: If (a)--(c) hold, ${\cal S}$ is consistent with the Markov sum rule.

\noindent {\em Proof}: It is easy to see that ${\cal S}_n$ is consistent with the Markov sum rule.
Hence we only need to check it for real parents of size $\geq n+1$. Such parents must be 
$C_{j,n}$-timid for some $j$.  As discussed before, Lemma A2 implies that its children
are either $C_{j,n}$-timid or not timid with respect to any $C_{k,n}, k=1..P$.
(b) ensures that 
the latter do not contribute to
the sum rule  and hence (c) ensures that the Markov sum rule is obeyed.

\vspace{3mm}

\noindent {\em Claim 4}: If (a)--(c) hold, ${\cal S}$ satisfies Bell causality.\\
\noindent {\em Proof}: 
Clearly ${\cal S}_n$ to stage $n-1$ is consistent with b.c.\ since it is just
the dynamical law of eqns (\ref{7}) or (\ref{12}). So we need to check b.c.\ with regard to real parents of size $\geq n$.\\

\noindent Case A) Real parents of size $n$: The only such causets are
$C_{j,n}$, each of which has a single real child. Let 
$C_1$ and $C_2$ be children of $C_{j,n}$ for some $j$. There are 2 cases:\\
\noindent (i) $C_1$ is timid and $C_2$ is virtual: Clearly there is no b.c.\ with offspring of 
any smaller causet $D\subset C_{j,n}$.  
Let $D\supset C_{j,n}$ be some real parent with children $D_1$ and $D_2$ such that 
$D_1,D_2$ enjoy a b.c.\ relation with $C_1,C_2$ (our notation is such that $D_1$ has
$C_{j,n}$ as its precursor set). Since $C_2$ is an ancestor of $D_2$, Claim 2 ensures that 
$D_2$ is virtual. As can be checked, this fact ensures that b.c.\ is satisfied. \\
\noindent (ii) $C_1$ and $C_2$ are virtual: b.c.\ is only non-trivial for offspring of
$D\supset C_{j,n}$.  Since $C_1,C_2$ are ancestors of $D_1,D_2$ respectively, Claim 2 ensures that $D_1,D_2$ 
are virtual and b.c.\ is satisfied.

\noindent Case B) Real parents of size $>n$:
The only real parents of size $>n$ are $C_{j,n}$-timid. Fix $j$ and let $C$ be a real $C_{j,n}$-timid
parent with children $C_1$ and $C_2$. There are 3 cases:\\
\noindent (i) $C_1$ is $C_{j,n}$-timid, and $C_2$ is not $C_{j,n}$-timid: Note that $C_2$ cannot be 
$C_{k,n}$-timid for any $k=1..P$ by Lemma A2. Hence $C_2$ is virtual. 
Obviously the 
union of precursor sets is either $C_{j,n}$ itself or is $C_{j,n}$-timid.  In
the former case it is easy to check that (a) ensures that b.c.\ is satisfied.  
In the latter case $C_1$ and $C_2$ are in a b.c.\ relation with 
appropriate children $D_1$ and $D_2$ of 
some real causet $D$ which contains $C_{j,n}$ as a partial stem.  From Lemma A3, since $D$ is real,
$D_1$ is $C_{j,n}$-timid and $D_2$ is not $C_{j,n}$-timid and hence
virtual. For $D\subset C$, g.c.\  (i.e. Claim 2)
ensures $D_1$ is real if $C_1$ is real and hence b.c.\ is satisfied. (If
$C_1$ is virtual then b.c.\ (iv) is an empty condition.)
If $D\supset C$, then 
$D_2$ is virtual (as argued above).
Further, since $C_1$ is an ancestor of $D_1$ in our definition of b.c.\ , Claim 2 ensures that
$D_1$ is virtual if $C_1$ is virtual and that $C_1$ is real if $D_1$ is real. Thus
b.c.\ holds. \\
\noindent (ii) $C_1$ and $C_2$ are not $C_{j,n}$-timid: Thus $C_1$ and $C_2$ are virtual and 
b.c.\ says nothing about transitions from $D\subset C$. For $D\supset C$, $D_1\supset C_1$ and
$D_2\supset C_2$ so that g.c.\  (i.e. Claim 2) ensures that 
$D_1, D_2$ are virtual and hence b.c.\ holds.\\
\noindent (iii) $C_1$ and $C_2$ are $C_{j,n}$-timid: b.c.\ is with children of $D$ which
contain a $C_{j,n}$-timid precursor set as a partial stem. Lemma A3 implies that $D$ and its
children $D_1, D_2$ are also $C_{j,n}$-timid.\footnote{In the degenerate case
where $D_1=D_2$ are the timid child of $C_{j,n}$, and $D$ is $C_{j,n}$
itself, the conclusion remains, since these are all in ${\cal P}_{C_{j,n}}$.}
Note that Lemma A2 implies that
if a causet is
$C_{j,n}$-timid it cannot be $C_{k,n}$-timid for $k\neq j$.  Thus b.c.\ holds by (c).

From the above it follows that if (a)--(c) hold then ${\cal S}$ is consistent
with b.c. 
This completes the proof of Lemma 4B.

\vspace{5mm}

\noindent {\bf Remark 1}: Any dynamical law relative to 
${\cal P}_C$  for some causet $C$, which satisfies the physical principles of section II 
relative to ${\cal P}_C$ 
is in correspondence with a growth process satisfying b.c., g.c.\ and the Markov rule and  which starts from 
the empty causet.  The correspondence is  that every causet $C^{\prime}$ of the latter defines
the causet $C^{\prime \prime}=C\cup C^{\prime}$ of the former, with every element of $C^{\prime}$ being to the future
of $C$ in $C^{\prime \prime}$. This remark is easily verified  by inspection.

\vspace{5mm}

\noindent{\bf The Final Picture}:
An iteration of Lemma 4 and Remark 1 yields the following picture. The formation of any real causet is
through a series of growth phases, each of whose transition probabilities are
given by eqns.\ (\ref{7}) or (\ref{12}). A
real  causet formed at the end of such a
growth stage 
will be called a {\em branch point causet}. Such a causet heralds a new phase 
of growth, with new values of the coupling constants. The transition
probabilities in this new stage are given by the same formulas (\ref{7}) or
(\ref{12}), with a completely new set of coupling constants $q_n$ or $t_n$
(these  can be freely chosen; for example, we could choose them to depend on the previous branch point 
causet in some way),
and the $\varpi$ and $m$ are interpreted ignoring the presence of the
previous branch point causet.
Every branch point causet $C_{\{(j_k,n_k),(j_{k-1},n_{k-1})..(j_1,n_1)\}}$
is labeled by a set of ordered pairs of natural numbers
$\{(j_k,n_k),(j_{k-1},n_{k-1})..(j_1,n_1)\}$.

The notation signifies that  $C_{\{(j_k,n_k),(j_{k-1},n_{k-1})..(j_1,n_1)\}}$ grew from the empty causet as 
a result of $k$ phases of independent growth.  
Call the causets which arise from each phase of independent sequential
growth as \emph{turtles}.  (i.e.\ a turtle is the difference between adjacent
branch point causets.)
The first stage of growth, characterized by $q_0 =1, q_i >0, i=1..n_1-1$,
resulted in a set of real turtles $C_{(j_1,n_1)}$ each of size $n_1$. Next, $q_{n_1}$ vanished. Choosing a fixed
$j_1$, the growth of the second set of turtles commenced, each new point
being to the future of the entirety of
$C_{(j_1,n_1)}$, and hence in correspondence with a sequential growth in which $C_{(j_1,n_1)}$ is replaced by the empty set.
The second set of turtles were fully formed when the effective $n_2$ antichain to $n_2+1$-antichain transition probability,
$q^{(j_1,n_1)}_{n_2}$, vanished. The third set of turtles' growth commenced from the parent $C_{\{(j_2,n_2),(j_1, n_1)\}}$
such that every new element was added to the future of $C_{\{(j_2,n_2),(j_1, n_1)\}}$ and so on.
Thus any real causet formed in the growth process consists of turtles stacked
one on top of another (see footnote 2).

\section{Conclusions}
The generic class of classical sequential growth dynamics derived in ref.\ \cite{RS} excluded the possibility that any of the transition probabilities vanished.
It is of interest to know if the picture changes drastically when zero transition probabilities are allowed.
In this work we 
generalized the considerations of ref.\ \cite{RS} 
to the case where the transition probabilities 
could vanish, deriving
the most general dynamical law which satisfied our generalizations of the 
physical principles of ref.\ \cite{RS}.
We found that this dynamics is
similar to that of ref\ \cite{RS}.
The transition probabilities are given by the same equations, except
 that the free parameters which define the dynamics are now allowed to
 vanish.  A unique feature emerges, however, when certain of these free
 parameters vanish.  In this case the development of the universe can
 abruptly change over to one with completely new values for these
 parameters, such that each element of the newly growing universe is
 born to the future of the entirety of the old universe.
In more technical terms, we found that our dynamics differs from that
of ref.\ \cite{RS} in two ways. 
One minor difference is that the coupling constants $t_n$ are allowed to
vanish, with the corresponding implications on the transition probabilities.
The more major difference arises from the vanishing of the `gregarious'
transition probabilities $q_n$.  Each such vanishing heralds the onset of a
new era, dubbed a `turtle', in which a completely new collection of coupling
constants $t_n$ may be used to describe the subsequent sequential growth, and
the added condition that each new element that arises
is to the future of the entirety of all the previous turtles.

It is interesting to note that, in spite of the fact that a turtle must 
finish its development at a particular stage in the growth
process, it is still compatible with covariance.  This is reminiscent of the
situation with cosmic renormalization and the formation of posts \cite{cosmic_renormalization}, in that the
existence of a post also implies a $C$-timid future evolution (where $C$ is a
causet with a unique maximal element).
A key
difference however is that cosmic renormalization occurs `within a single
turtle', with a single set of coupling constants.
One can show that originary dynamics (e.g.\ that which occurs after a post) can
be identified with a turtle dynamics in which the `origin element' is a
turtle followed by an infinite second turtle.\footnote{One recovers the originary dynamics described in
ref.\ \cite{cosmic_renormalization} by choosing the coupling constants of the
second turtle $\widetilde{t_n}$ in terms of those of the originary evolution
$t_n$ by
$\widetilde{t_0}=1, \widetilde{t_n}=\frac{t_n+t_{n+1}}{t_1}$.}
When we consider something like a `double post', in which $C$ has
multiple maximal elements, it is important to note that only the restriction
to $C$-timid causets is compatible with g.c.  Causets which attempt to
generalize originary dynamics by enforcing a causet with more than one maximal
element as a full stem (as suggested in ref.\ \cite{RS}) will violate covariance.


As emphasized in the main body of the paper, the relevant output of any physically satisfactory dynamical law 
is the specific probability  of formation of any causet rather than the transition probabilities of 
individual transitions. In the case of exclusively non-vanishing transition probabilities, one can be derived
from the other, but in the case of possibly vanishing transition probabilities, two dynamical laws differing in their 
assignations (consistent
with the physical principles discussed in section II) of transition probabilities to links emanating from
virtual causets still yield the same specific probabilities of
formation. Though indeed even the specific probabilities
of formation themselves are not physically relevant since they pertain to finite causets formed at a
particular stage of the growth process.  What is of physical relevance is the resulting probability measure on a suitable space
of completed, unlabeled causets \cite{RSmeasure}, or rather sets of such
causets to which can be attached a physical meaning. The relevant analysis for the dynamics of ref.\ \cite{RS}, 
showing that indeed such a measure 
can be defined and characterized in terms of answers to physical questions, has been done in ref.\ \cite{RSmeasure}.
The corresponding analysis for the dynamics described in this paper has recently been completed by Dowker and 
Surya \cite{sumati}.
The key open issue is of course the formulation of the quantum dynamics. It is hoped that our understanding of
the most general classical dynamics, in conjunction with the work of 
ref.\ \cite{sumati}, may be of some use in the formulation of the quantum dynamics and 
in investigating aspects of its (semi)classical limit.

\noindent{\bf Acknowledgments}: MV would like to thank Rafael Sorkin for suggesting this problem
and for his warm hospitality and time. 
MV would also like to thank Rafael Sorkin and 
Graham Brightwell for their generosity with regard to authorship issues and 
Graham Brightwell specially for sharing his insights.

DR would like to thank the Williams Watrous Couper Fund for generous travel
support, the Raman Research Institute for kind hospitality wherein this work
was started, and S. Varadarajan for crucial assistance.


\vspace{5mm}

\noindent{\bf Appendix}

\noindent{\bf Lemma A1}:
Given any dynamical law consistent with the physical principles of section II, we may set the transition probabilities to zero
for any links emanating from any virtual causet. The new dynamical law thus defined is also consistent with the principles of section II
and is equivalent to the original one in that it  provides the same specific probabilities of formation.

\noindent{\bf Proof}: The proof is trivial since (a) b.c.\ and the Markov sum rule apply only to real parents, (b) virtual/real
causets of the original growth  remain virtual/real in the new one, and (c) transition probabilities for links along paths
leading from the empty causet to any real causet are unchanged.

\vspace{5mm}

\noindent{\bf Lemma A2}:  Let $C$ be a causet of cardinality $n$. Then every path from the empty set to any $C$-timid causet 
must pass through $C$ at stage $n$.

\noindent{\bf Proof} (by contradiction): Suppose there is a path from the empty set to a $C$-timid causet passing through $C^{\prime}$ at 
stage $n$, with $C^{\prime}$ not isomorphic to $C$. Since the final causet is $C$-timid, $C$ must form at some stage 
$p > n$. But then there will  be $p-n$ elements not in $C$ which are not to
the future of every maximal element of $C$.  Hence no such path can exist.

{}From Lemma A2 it immediately follows that:\\
\noindent (i) No path exists from a $C$-timid causet to a $C^{\prime}$-timid causet where $C$ and $C^{\prime}$ are non-isomorphic
causets of cardinality $n$.\\
\noindent (ii) No path exists from a causet which is not timid with respect to $C$ and which has cardinality $>n$ 
to a $C$-timid causet since,
as can easily be verified, the former can be formed along a path not passing through $C$ at stage $n$.

\vspace{5mm}

\noindent{\bf Lemma A3}: If a causet $C$ is $C^{\prime}$-timid and has partial stem $C^{\prime \prime}$ with 
$C^{\prime}$ and $C^{\prime \prime}$ both of size $n$, then
$C^{\prime}=C^{\prime \prime}$. 

\noindent{\bf Proof} (by contradiction): Suppose $C^{\prime}\neq C^{\prime \prime}$ in $C$. Then
there exists $x\in C \setminus C^{\prime}$ such that 
$x\in C^{\prime \prime}$. Since $C$ is  $C^{\prime}$-timid, $\mathrm{past}(x) \supseteq C^{\prime}$. Since $C^{\prime \prime}$
is a partial stem,  $\mathrm{past}(x)\subseteq C^{\prime \prime}$. This implies that $C^{\prime \prime}\supseteq C^{\prime}$
which in turn implies that $C^{\prime \prime}=C^{\prime}$ since they are of the same cardinality.

\end{document}